\documentclass[preprint,preprintnumbers,amsmath,amssymb]{revtex4}

\usepackage{graphicx,color}
\usepackage{bm}

\begin{document}

\newcommand{\AR}[1]{{{\bf AR:} \em #1}}
\newcommand{\CP}[1]{{{\bf CP:} \em #1}}
\newcommand{\OLD}[1]{{\tiny #1}}

\newcommand{\mfs}{Mn$_{1-x}$Fe$_{x}$Si}
\newcommand{\mcs}{Mn$_{1-x}$Co$_{x}$Si}
\newcommand{\fcs}{Fe$_{1-x}$Co$_{x}$Si}
\newcommand{\cso}{Cu$_{2}$OSeO$_{3}$}

\newcommand{\rxx}{$\rho_{xx}$}
\newcommand{\rxy}{$\rho_{xy}$}
\newcommand{\rxytop}{$\rho_{\rm xy}^{\rm top}$}
\newcommand{\Drxyt}{$\Delta\rho_{\rm xy}^{\rm top}$}
\newcommand{\Sxy}{$\sigma_{xy}$}
\newcommand{\Sxya}{$\sigma_{xy}^A$}

\newcommand{\bco}{$B_{\rm c1}$}
\newcommand{\bct}{$B_{\rm c2}$}
\newcommand{\bao}{$B_{\rm A1}$}
\newcommand{\bat}{$B_{\rm A2}$}
\newcommand{\beff}{$B_{\rm eff}$}
\newcommand{\bmm}{$B_{\rm m}$}
\newcommand{\bs}{$B_{\rm S}$}
\newcommand{\btop}{$B_{\rm top}$}
\newcommand{\bnfl}{$B_{\rm NFL}$}

\newcommand{\tc}{$T_{\rm c}$}
\newcommand{\tmax}{$T_{\rm max}$}
\newcommand{\ts}{$T_{\rm S}$}

\newcommand{\pc}{$p_{\rm c}$}

\newcommand{\mb}{$\mu_0\,M/B$}
\newcommand{\dmdb}{$\mu_0\,\mathrm{d}M/\mathrm{d}B$}
\newcommand{\ddmddb}{$\mathrm{\mu_0\Delta}M/\mathrm{\Delta}B$}
\newcommand{\cm}{$\chi_{\rm M}$}
\newcommand{\cac}{$\chi_{\rm ac}$}
\newcommand{\rechi}{${\rm Re}\,\chi_{\rm ac}$}
\newcommand{\imchi}{${\rm Im}\,\chi_{\rm ac}$}

\newcommand{\ozz}{$\langle100\rangle$}
\newcommand{\ooz}{$\langle110\rangle$}
\newcommand{\ooo}{$\langle111\rangle$}
\newcommand{\too}{$\langle211\rangle$}

\renewcommand{\floatpagefraction}{0.5}


\title{Formation of a topological non-Fermi liquid in MnSi}

\author{R. Ritz}
\affiliation{Physik Department E21, Technische Universit\"at M\"unchen, D-85748 Garching, Germany}

\author{M. Halder}
\affiliation{Physik Department E21, Technische Universit\"at M\"unchen, D-85748 Garching, Germany}

\author{M. Wagner}
\affiliation{Physik Department E21, Technische Universit\"at M\"unchen, D-85748 Garching, Germany}

\author{C. Franz}
\affiliation{Physik Department E21, Technische Universit\"at M\"unchen, D-85748 Garching, Germany}

\author{A. Bauer}
\affiliation{Physik Department E21, Technische Universit\"at M\"unchen, D-85748 Garching, Germany}

\author{C. Pfleiderer}
\affiliation{Physik Department E21, Technische Universit\"at M\"unchen, D-85748 Garching, Germany}

\date{\today}

\maketitle

\textbf{Fermi liquid theory provides a remarkably powerful framework for the description of the conduction electrons in metals and their ordering phenomena, such as superconductivity, ferromagnetism, and spin- and charge-density-wave order.  A different class of ordering phenomena of great interest concerns spin configurations that are topologically protected, that is, their topology can be destroyed only by forcing the average magnetization locally to zero \cite{Chaikin}. Examples of such configurations are hedgehogs (points at which all spins are either pointing inwards or outwards) or vortices. A central question concerns the nature of the metallic state in the presence of such topologically distinct spin textures. Here we report a high-pressure study of the metallic state at the border of the skyrmion lattice in MnSi, which represents a new form of magnetic order composed of topologically non-trivial vortices \cite{Muehlbauer:Science2009}. When long-range magnetic order is suppressed under pressure, the key characteristic of the skyrmion lattice -- that is, the topological Hall signal due to the emergent magnetic flux associated with their topological winding -- is unaffected in sign or magnitude and becomes an important characteristic of the metallic state. The regime of the topological Hall signal in temperature, pressure and magnetic field coincides thereby with the exceptionally extended regime of a pronounced non-Fermi-liquid resistivity \cite{Pfleiderer:Nature2001, Doiron:Nature03}. The observation of this topological Hall signal in the regime of the NFL resistivity suggests empirically that spin correlations with non-trivial topological character may drive a breakdown of Fermi liquid theory in pure metals.} 

To address the nature of the metallic state at the border of long-range topological order we have selected the  B20 compound MnSi. At zero pressure, $p=0$, MnSi undergoes a fluctuation-induced first-order transition to helimagnetic order at $T_c=29.5\,{\rm K}$ \cite{Janoschek:PRB2013}. The helimagnetism originates in a hierarchy of three energy scales, comprising ferromagnetic exchange on the strongest scale, Dzyaloshinsky-Moriya interactions on an intermediate scale and higher-order spin-orbit coupling on the weakest scale \cite{Landau}. As a function of field, $B$, conical order appears for $B>B_{\rm c1}\approx0.1\,{\rm T}$, and this is followed by a spin-polarized state for $B>B_{\rm c2}\approx0.6\,{\rm T}$. A phase pocket in the vicinity of $T_c$, known as the A-phase, supports the skyrmion lattice \cite{Muehlbauer:Science2009}. The magnetic phase diagram of MnSi including the skyrmion lattice is thereby generic for all helimagnetic B20 compounds, regardless whether they are high-purity metals \cite{Muehlbauer:Science2009,Yu:NatureMaterials2011}, semiconductors \cite{Muenzer:PRB2010,Yu:Nature2010} or insulators \cite{Seki:Science2012,Adams:PRL2012}. 

As a function of pressure, the helimagnetic transition in MnSi vanishes above $p_c\approx14.6\,{\rm kbar}$ without quantum criticality \cite{Pfleiderer:Nature2004,Pfleiderer:Science07,Uemura:NaturePhysics07}. Yet the resistivity changes from the $T^2$ dependence of a Fermi liquid to the $T^{3/2}$ dependence of a non-Fermi liquid (NFL) when $p$ exceeds $p_c$. The exceptionally wide NFL range \cite{Pfleiderer:Nature2001,Doiron:Nature03} and the lack of sample dependence of the $T^{3/2}$ coefficient contrast with the excellent quantitative description of MnSi as a weak itinerant magnet \cite{Lonzarich:JPSS1985}, suggesting that the cause of the NFL behaviour may be an intrinsic mechanism mimicking the effects of disorder and glassiness. This notion is supported by neutron scattering, NMR and muon spin resonance measurements, suggesting partial magnetic order on timescales between $10^{-10}\,{\rm s}$ and $10^{-11}\,{\rm s}$ \cite{Pfleiderer:Nature2004,Uemura:NaturePhysics07}. In turn, several theoretical studies \cite{Tewari:PRL2006,Binz:PRL2006,Roessler:Nature2006} explored a proliferation of topological spin textures as the cause of the partial order, with a possible link to the NFL resistivity \cite{Kirkpatrick:PRL2010}. However, until now, evidence for topologically non-trivial spin textures in the NFL regime as well as a link between such textures and the NFL behaviour has not been reported.

A unique experimental probe of the topology of magnetically ordered states is the Hall effect, which reflects the Berry phases developed by the electrons as they follow the spin orientation of the magnetic structure. The consequences of these Berry phases may be viewed in terms of an emergent magnetic field with two limiting types of behaviour \cite{Ritz:PRB2013}. For variations on atomic scales, the emergent field acts essentially in reciprocal space, giving rise to dissipationless Hall currents; this is referred to as the intrinsic anomalous Hall effect. In contrast, for smooth variations on length scales much larger than the Fermi wavelength, the emergent field acts similarly to a real magnetic field, giving rise to a topological Hall signal. Experimentally, these two limits may be readily distinguished in terms of their dependence on temperature and magnetization. Namely, the Hall resistivity associated with the intrinsic anomalous Hall contribution scales with the square of the longitudinal resistivity, $\rho_{\rm xx}^2$. Therefore, in ``good'' metals, where $\rho_{xx}$ rapidly decreases with temperature, the intrinsic anomalous contribution to $\rho_{xy}$ also decreases with decreasing temperature. This is different from the topological contribution to $\rho_{\rm xy}$, which increases rapidly with decreasing temperature, owing to the increase in the spin polarization and a reduction of spin-flip scattering \cite{Ritz:PRB2013}. 

At ambient pressure, the Hall effect in MnSi is dominated by the sum of a normal contribution and an intrinsic anomalous contribution \cite{Lee:PRB07}. In addition, a topological Hall signal has been observed in the skyrmion lattice phase \cite{Neubauer:PRL2009,Ritz:PRB2013}. Moreover, a giant topological Hall signal exists for pressures between 6 and 12\,kbar \cite{Lee:PRL09}. However, that study \cite{Lee:PRL09} failed to connect the giant topological Hall signal experimentally with the skyrmion lattice phase at ambient pressure and the NFL behaviour at high pressure. In addition, the size and field range of the Hall signal seemed anomalously large and at odds with the behaviour expected at ambient pressure. 

To search for a link between the skyrmion lattice and the NFL resistivity, and to resolve the origin of the giant topological Hall signal reported in Ref.\,\cite{Lee:PRL09}, we performed an extensive high-pressure study exploring the roles of sample purity, pressure transmitter, cooling procedure and sample orientation. To this end, we assembled eight pressure cells covering nearly 40 pressure points (details of the experimental methods are reported in Ref.\,\cite{Ritz:PRB2013} and Supplementary Information). As an important first step in studying the complex phase diagram near $p_c$, we demonstrated that the giant topological Hall signal at low pressures reported in Ref.\,\cite{Lee:PRL09} is connected to the skyrmion lattice phase at $p=0$. We found that the large size of the signal is intrinsic, and that the large field range is most probably due to low sample purity and inhomogeneous pressure conditions.

Here we report that the topological Hall signal of the skyrmion lattice evolves into a prominent characteristic of the entire NFL regime for pressures up to $18.1\,{\rm kbar}$, the highest pressure we investigated, which is much greater than $p_c$. Our conclusions are based on the sign and size of the topological Hall signal and the observation of clear boundaries of this signal (Fig.\,\ref{figure1}\,a, b) that reproduce the boundaries of the NFL behaviour of {\rxx} as a function of pressure, magnetic field and temperature \cite{Pfleiderer:Nature2001,Doiron:Nature03}. First, for pressures exceeding $p^*\approx12\,{\rm kbar}$ there is a clear crossover with decreasing temperature from the regime at high temperatures to a NFL resistivity at $T^* \approx 12\,{\rm K}$, and {\rxx} settles into a stable $T^{3/2}$ NFL dependence below $\approx 8\,{\rm K}$. In fact, for what follows below it is essential to emphasize that for $T_c\lesssim T^*$, the crossover and associated NFL behaviour emerges between $T_c$ and $\approx T^*$ (Ref.\,\cite{Pfleiderer:Nature2001}; below $T_c$ Fermi liquid behaviour is observed). Second, as mentioned above, the dependence of {\rxx} on $T$, $\rho_{\rm xx}(T) \propto T^{\alpha}$, changes abruptly at {\pc} from a Fermi liquid form ($\alpha=2$), to the NFL form ($\alpha=3/2$) (Fig.\,\ref{figure1}\,a, inset). Third, for $p>p_c$ the NFL behaviour returns abruptly as a function of field to Fermi liquid behaviour at a value, {\bnfl}, that coincides with the itinerant metamagnetic transition, at {\bmm} (Fig.\,\ref{figure1}\,b, inset; we refrain here and in Fig.\,\ref{figure2}\,b from showing exponents in the centre of the crossover very close to {\bnfl}). 

To be able to present our main observations further below, which requires distinguishing topological contributions to the Hall signal from those that are anomalous, we show in Fig.\,\ref{figure1}\,c, d our magnetotransport data over a large field range, $-14$ to $14\,{\rm T}$. At 2.8\,K and at low pressures, {\rxx} drops by $\sim10\,\%$ at {\bct}, an this is followed by a gradual increase towards high fields (Fig.\,\ref{figure1}\,c, upper panel). Above $T_c$, this decrease in {\rxx} broadens and shifts to higher fields (Fig.\,\ref{figure1}\,c, lower panel). For $p>p_c$, {\rxx} displays a similar decrease at {\bmm} (the itinerant metamagnetic transition) with a temperature dependence for {\bmm}  that is empirically similar to what is seen above $T_c$ for low pressures, that is, {\bmm} shifts to higher values. This is consistent with spin scattering processes in the NFL regime that are quenched above {\bmm}. Furthermore, the Hall signal {\rxy} at 2.8\,K is essentially featureless and linear over the range $-14$ to $14\,{\rm T}$ for all pressures studied (Fig.\,\ref{figure1}\,d, upper panel), with the exception of the topological contributions at low fields, which we address below. At higher temperatures, a ``knee-shaped'' anomalous Hall contribution develops in addition to the signal seen at 2.8\,K (Fig.\,\ref{figure1}\,d, lower panel). This includes, at higher pressures, the emergence of an additional ``shoulder'' and related maximum at a field {\bs}, which reaches several Tesla and coincides with the itinerant metamagnetism at {\bmm} at low temperatures and high pressures, as inferred from the a.c. susceptibility \cite{Thessieu:JPCM97,Ritz:PRB2013}. A discussion of the associated temperature dependences is presented in Supplementary Information. Hence, for large fields {\rxy} is dominated by the sum of a normal contribution and an intrinsic anomalous contribution \cite{Lee:PRB07}, of which the latter displays an additional shoulder (see also Fig.\,\ref{figure3}). The shoulder in {\rxy} can therefore be accounted for neither by variations in {\rxx} (entering the Hall conductivity $\sigma_{\rm xy} = -\rho_{\rm xy} / \left( \rho_{\rm xy}^2 + \rho_{\rm xx}^2 \right)$, which is proportional to the intrinsic anomalous Hall effect) nor by the magnetization, the field dependence of which is qualitatively similar above $T_c$ for all pressures, with no signs of a shoulder \cite{Thessieu:JPCM97,Ritz:PRB2013}. This suggests a more subtle connection between the shoulder in {\rxy} and the itinerant metamagnetism, but this connection is not important for the conclusions of our study.

Typical data highlighting the topological contributions to the Hall signal for samples with high residual resistivity ratios (RRRs) are summarized in Fig.\,\ref{figure2}. Qualitatively similar data for samples with low RRRs are presented in Supplementary Information. To illustrate the relationship between the topological Hall signal and the nature of the metallic state inferred from the temperature dependences of {\rxx}, we use the following colour scheme: white for the paramagnetic regime, sky blue for the Fermi liquid and orange for the NFL. For low pressures, a topological Hall signal arises exclusively in the A-phase \cite{Ritz:PRB2013} (red shading). It is important to emphasize that the change in slope of {\rxy} in this regime (for example at 0.5\,T for 7\,kbar and 16.6\,K) is due to the conical-to-ferromagnetic transition at {\bct} and is therefore completely unrelated the itinerant metamagnetism at {\bs} at high pressures and above $T_c$.

As $T_c$ decreases below $T^* \approx 12\,{\rm K}$, the topological signal is also present above $T_c$ up to $\sim T^*$ and over a wide field range where the magnitude and sign of the signal are unchanged compared with the data below $T_c$ (see data for 13.7\,kbar in Fig.\,\ref{figure2}\,a). Because the shoulder in the itinerant metamagnetism data at these temperatures and pressures is located at a field strength of around several Teslas (well outside the field range shown here; see Supplementary Fig.\,4), this unambiguously connects the topological Hall signal in the A-phase with the NFL resistivity. At pressures greater than $p_c$, the topological contribution is observed across the entire NFL regime. We have confirmed this observation in several samples of differing purity, where the topological Hall contribution tends to be larger and broadened for these samples with lower RRRs. We thereby note that, regardless of the RRR, there are no hints of metastable behaviour such as that observed under field cooling at low pressures (see Supplementary Fig.\,6 and fig.\,7 in Ref.\,\cite{Ritz:PRB2013}).

To justify the positioning of regime boundaries in Fig.\,\ref{figure2}\,a, we show in Fig.\,\ref{figure2}\,b, for $p=14.7\,{\rm kbar}>p_c$, a typical comparison of {\rxx} with the Hall resistivity after subtracting the normal Hall contributions, $\rho_{\rm xy}-\rho_{\rm xy,norm}$ (where $\rho_{\rm xy,norm}$ was inferred from data above {\bct}), and with the exponent, $\alpha$, of the temperature dependence of the resistivity at selected fixed fields. As a function of field, the following characteristic features coincide at {\bnfl}: the onset of the decrease in the magnetoresistance suggestive of a field-induced suppression of a magnetic scattering mechanism, the disappearance of the topological Hall contribution (Fig.\,\ref{figure2}\,b, red shading) and the change from Fermi liquid to NFL resistivity. We can therefore confirm that the field dependence of {\rxx} neither qualitatively nor quantitatively causes variations of the intrinsic anomalous Hall contribution ($\sigma_{\rm xy}\approx -\rho_{\rm xy}/\rho_{xx}^2$) that may be mistaken for a topological Hall signal. Thus, the topological Hall signal and the NFL resistivity coincide as functions of $T$, $B$ and $p$.

In Fig.\,\ref{figure3}, we show the relevant regions of the pressure-temperature phase diagram and typical magnetic phase diagrams. For zero field, NFL behaviour emerges below $T^*$ and above $T_c$ as well as below $T^*$ for pressures greater than $p_c$. The NFL behaviour is accompanied by a topological Hall signal that is present even for very low fields, as marked by NFL+TH in each panel. For pressures below $p^* \approx 12\,{\rm kbar}$, the magnetic phase diagram remains qualitatively unchanged as compared with ambient pressure (Fig.\,\ref{figure3}\,b), and the topological Hall signal is enhanced as reported in Ref.\,\cite{Ritz:PRB2013}. Qualitatively new behaviour emerges for $p>p^*$, where the topological Hall contribution survives above $T_c$ and $p_c$. This is illustrated in Fig.\,\ref{figure3}\,c, d, e where the boundaries of the regime of the NFL resistivity are shown as a function of $T$, $B$ and $p$. 

The sign and the magnitude of the topological Hall signal arises from a combination of the strength of the emergent field, given by the topological winding number for the magnetic unit cell; the (local) spin polarization of the conduction electrons; and an average over individual bands taking into account different scattering processes \cite{Ritz:PRB2013}. The observation of neutron scattering intensity above $p_c$ at a wavelength which corresponds to that of the helical state at low pressures \cite{Pfleiderer:Nature2004,Pfleiderer:PRL07} implies that the continuous evolution of the topological Hall signal from the A-phase to the NFL regime must be closely connected to an unchanged topological winding number. As a caveat, neutron intensity is observed only in a small region of the NFL regime, whereas the topological Hall signal we report here is seen everywhere. Thus, the topological winding must be insensitive against fluctuations above $p_c$, at least on time scales relevant to the Hall effect. As a consequence, any of the following scenarios (or combinations thereof) can arise in the high-pressure state of MnSi: spontaneous formation of randomly oriented skyrmions at $B=0$ that are stratified in a magnetic field; formation of skyrmions even at very low fields; strong fluctuations between helical modulations and skyrmionic textures. It therefore seems likely that strong quantum fluctuations promote this state.

Our study identifies a topological Hall signal as a prominent characteristic of the NFL regime of MnSi. The sign and magnitude of the Hall signal indicate that the topological winding of skyrmions as seen at $p=0$ is the long-sought intrinsic mechanism causing the $T^{3/2}$ NFL dependence. This dependence is characteristic of a strong divergence of the quasiparticle self-energy as $T\to0$, which suggests a generic breakdown of Fermi liquid theory due to spin correlations with topologically non-trivial character. Notably, several theoretical studies suggest various mechanisms leading to the formation of non-trivial spin textures at the zero-temperature border of itinerant ferromagnetism for different reasons \cite{Belitz:PRL2002,Conduit:PRL2009}. In fact, experimentally a $T^{3/2}$ NFL resistivity has also been observed near putative ferromagnetic quantum phase transitions, for example in ZrZn$_2$ \cite{Smith:Nature2008} and Ni$_3$Al. However, if this is related to complex spin textures, their average topological winding would be degenerate and no topological Hall effect would therefore be expected. Hence, the topological character of the NFL regime we report here for MnSi may turn out to be the first example of a more general phenomenon, in which the full suppression of the magnetization is generically preceded by the formation of complex spin textures. For the case of MnSi, the peculiar stability of the NFL behaviour may thereby be inherited from the chiral character of the Dzyaloshinsky-Moriya spin-orbit interaction.

\newpage
\section*{Methods summary}

\textbf{Sample preparation.} Single crystals of MnSi were grown by optical float zoning under ultrahigh-vacuum-compatible conditions \cite{Neubauer:RSI2011}. The specific heat, susceptibility and resistivity of small pieces taken from these single crystals were in excellent agreement with established results. Samples with different RRRs were studied as summarized in Supplementary Information. Samples for the measurements reported here were oriented by Laue X-ray diffraction, cut with a wire saw and carefully polished to size. Current leads were soldered to the small faces of the sample, and platinum wires for the voltage pick-up were spot-welded onto the surface of the sample.

\textbf{Magnetotransport under pressure.} The Hall resistivity and longitudinal resistivity were measured simultaneously using a standard six-terminal phase-sensitive detection system. The voltage signals were amplified with impedance-matching transformers to optimize the signal-to-noise ratio. Low excitation frequencies and excitation currents were applied to minimize parasitic signal pick-up. High hydrostatic pressures were generated with a Cu:Be clamp cell using various different pressure transmitter as described in detail in Ref.\,\cite{Ritz:PRB2013} and Supplementary Information. The sample was suspended by the current and voltage leads, using a Teflon platform to fix the location of the wires and thereby the sample orientation. Data were recorded at temperatures down to 1.5\,K under magnetic fields in the range $-14$ to $14\,{\rm T}$, using a conventional superconducting magnet system. Further details may be found in Ref.\,\cite{Ritz:PRB2013} and Supplementary Information.

\newpage
\section*{Acknowledgements}

We wish to thank P. B\"oni, K. Everschor, M. Garst, M. Janoschek, S. Mayr and A. Rosch for discussions and support. R.R., M.H., A.B., M.W. and C.F. acknowledge financial support through the TUM Graduate School. Financial support through DFG TRR80 and FOR960 as well as ERC-AdG (291079 TOPFIT) are gratefully acknowledged. 

\section*{Author Contributions}
R.R. and C.P. developed the experimental set-up; 
R.R. performed the transport measurements; 
M.H. and M.W. performed magnetization measurements; 
C.F. wrote the software for analysing the data;
A.B. grew the single-crystal samples and characterized them; 
R.R. and C.P. analysed the experimental data;
C.P. supervised the experimental work; 
C.P. proposed this study and wrote the manuscript; 
all authors discussed the data and commented on the manuscript;
correspondence should be addressed to R.R. (robert.ritz@frm2.tum.de) or C.P. (christian.pfleiderer@frm2.tum.de).

\newpage


\newpage


\clearpage
\thispagestyle{empty}
\centerline{\includegraphics[width=16cm]{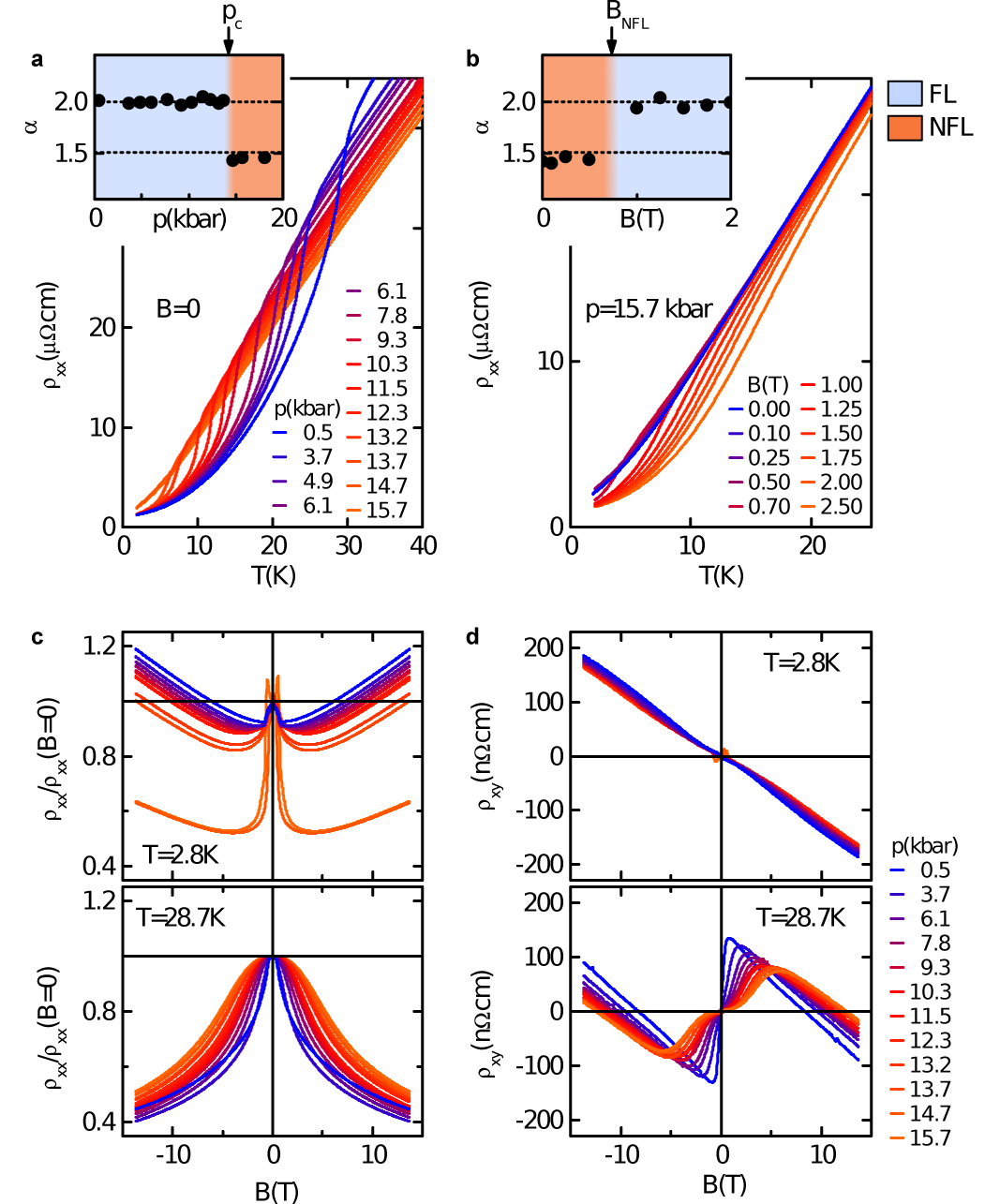}}
\bigskip
\bigskip
\Large\textbf{Fig.~1}
\clearpage
\begin{figure}[h]
\caption{\textbf{Temperature and field dependence of the resistivity, {\rxx}, and Hall resistivity, {\rxy}, of MnSi over a wide range.} Data shown here were measured in pressure pc8 (see also Ref.\,\cite{Ritz:PRB2013} and Supplementary Information). \textbf{a,} {\rxx} at various pressures (see also Ref.\,\cite{Pfleiderer:PRB97}). Inset, exponent $\alpha$ inferred from {\rxx} (see also Refs.\,\cite{Pfleiderer:Nature2001,Doiron:Nature03}). \textbf{b,} {\rxx} at various fields for $p=15.7\,{\rm kbar}$. Inset, exponent $\alpha$ at $p=15.7\,{\rm kbar}$. \textbf{c,} {\rxx} at 2.8\,K and 28.7\,K up to 15.7\,kbar. \textbf{d}, {\rxy} up to 15.7\,kbar. At 28.7\,K a strong knee-shaped intrinsic anomalous Hall contribution emerges; that is, an additional shoulder emerges on top of the intrinsic anomalous Hall contribution that merges with the itinerant metamagnetism above $p_c$ and at low temperatures. FL, Fermi liquid; NFL, non-Fermi liquid.}
\label{figure1}
\end{figure}

\clearpage
\thispagestyle{empty}
\centerline{\includegraphics[width=9.5cm]{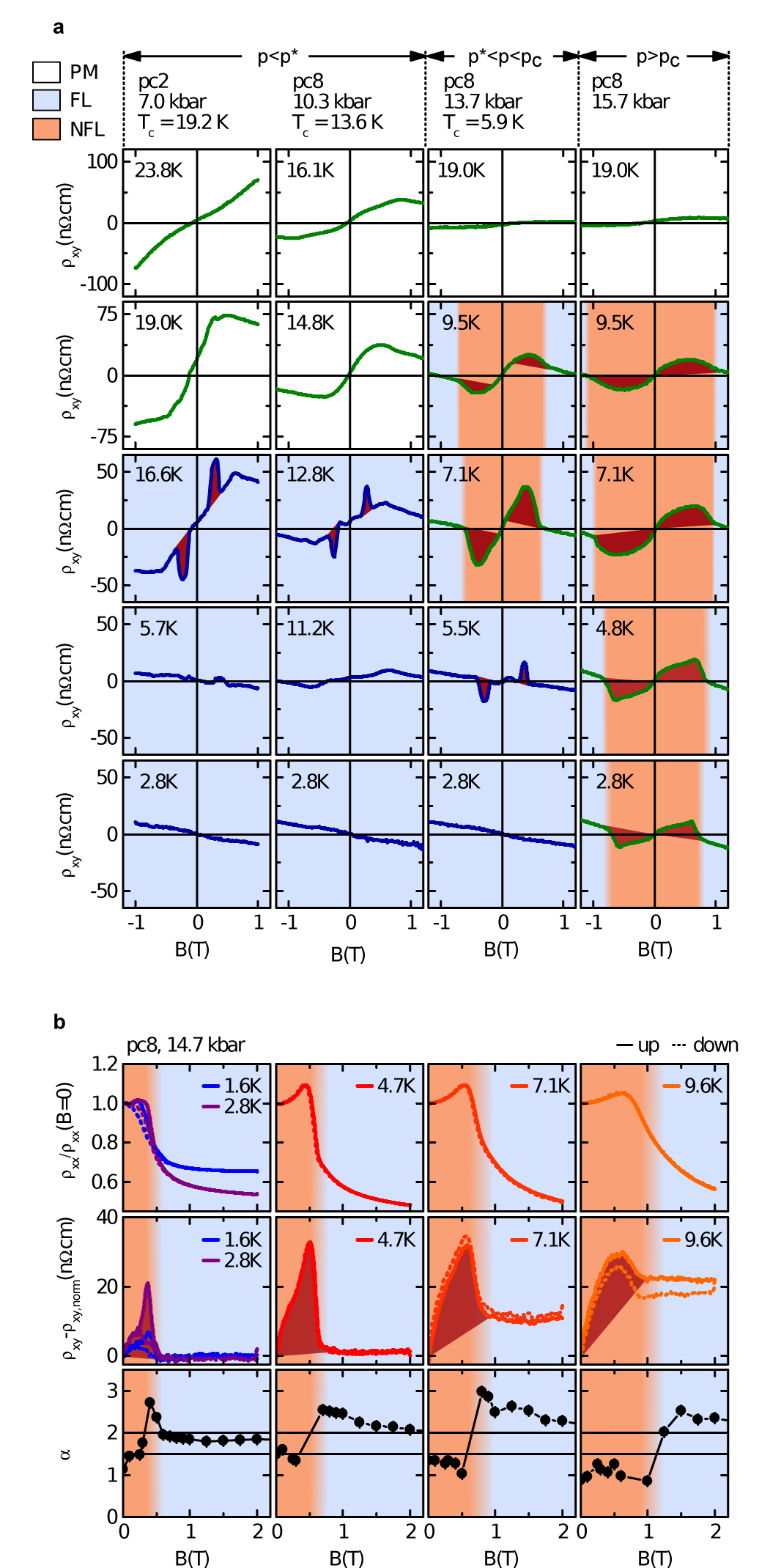}}
\bigskip
\bigskip
\Large\textbf{Fig.~2}
\clearpage
\begin{figure}[h]
\caption{\textbf{Hall resistivity, {\rxy}, and magnetoresistance, {\rxx}, for low fields at various pressures.} \textbf{a,} {\rxy} for a wide range of pressures and temperatures. \textbf{b,} Comparison of the magneto-resistance, {\rxx}; the Hall resistivity after subtraction of normal contributions, $\rho_{\rm xy}-\rho_{\rm xy,norm}$; and the exponent of the temperature dependence of the electrical resistivity, $\alpha$, at various temperatures, for a pressure greater than $p_c$. With increasing field, $\alpha$, changes from NFL behaviour ($\alpha\approx3/2$) to Fermi liquid behaviour ($\alpha\approx2$) at the same field value, above which the topological Hall contribution has vanished. PM, paramagnet; pc2 and pc8 refer to the pressure cells studied \cite{Ritz:PRB2013} (Supplementary Information).
} \label{figure2}
\end{figure}

\clearpage
\thispagestyle{empty}
\centerline{\includegraphics[width=9.0cm]{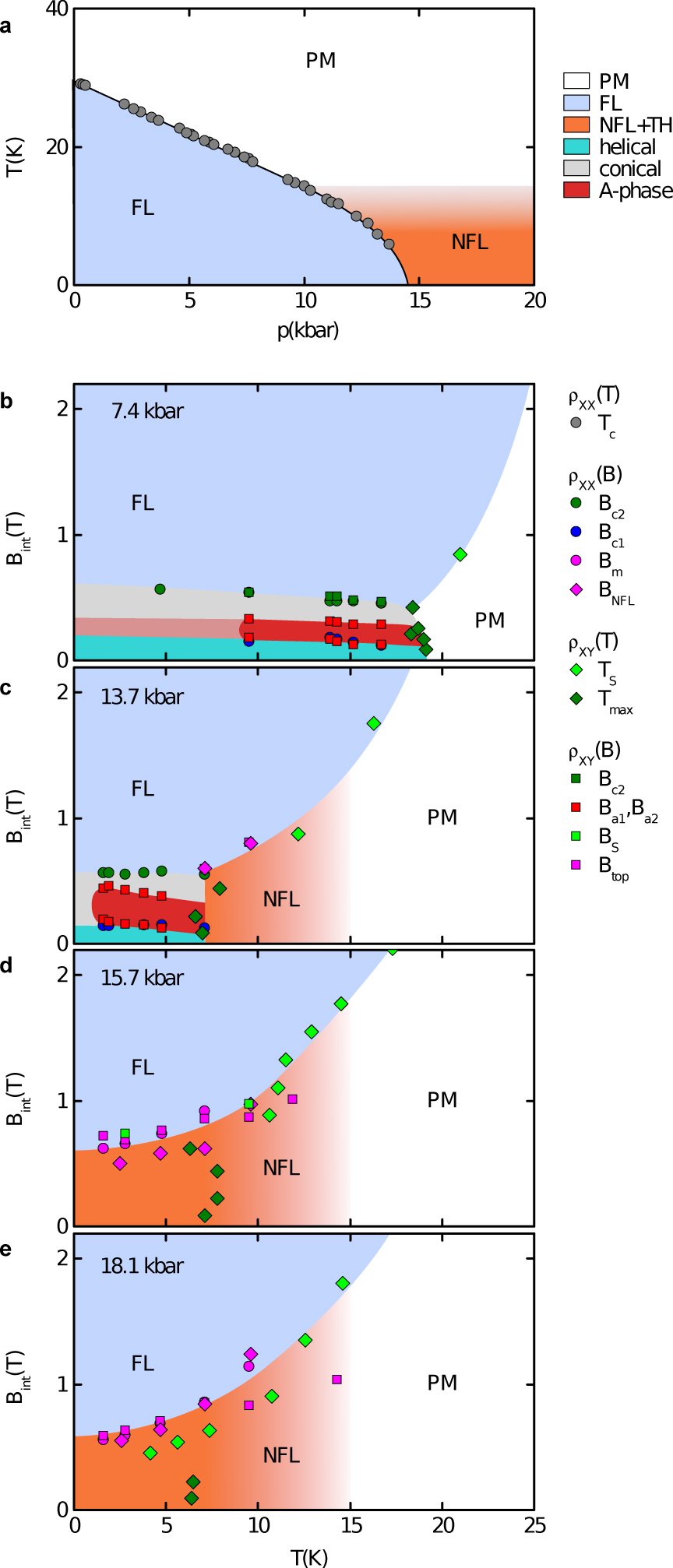}}
\bigskip
\bigskip
\Large\textbf{Fig.~3}
\clearpage
\begin{figure}[h]
\caption{\textbf{Phase diagrams of MnSi.} \textbf{a,} Temperature-pressure phase diagram. NFL+TH refers to the regime of an NFL resistivity in which a small field establishes a topological Hall (TH) signal. \textbf{b,} Magnetic phase diagram at $p=7.4\,{\rm kbar}$, displaying a strongly enhanced topological Hall signal in the A-phase as reported in Ref.\,\cite{Ritz:PRB2013}. \textbf{c,} Magnetic phase diagram at {\tc} and $p=13.7\,{\rm kbar}$. The transition at {\tc} is first order, and the onset of NFL resistivity occurs at a temperature above $T_c$. A topological Hall contribution is observed in the A-phase and the NFL regime above $T_c$. \textbf{d, e,} Magnetic phase diagrams at $p=15.7\,{\rm kbar}$ and $p=18.1\,{\rm kbar}$, above $p_c$. The field and temperature range of the topological Hall signal extents over the entire field and temperature range of the $T^{3/2}$ NFL resistivity. See Supplementary Information for illustrated definitions of all characteristic fields and temperatures.} \label{figure3}
\end{figure}


\end{document}


\newcommand{\AR}[1]{{{\bf AR:} \em #1}}
\newcommand{\CP}[1]{{{\bf CP:} \em #1}}
\newcommand{\OLD}[1]{{\tiny #1}}

\newcommand{\mfs}{Mn$_{1-x}$Fe$_{x}$Si}
\newcommand{\mcs}{Mn$_{1-x}$Co$_{x}$Si}
\newcommand{\fcs}{Fe$_{1-x}$Co$_{x}$Si}
\newcommand{\cso}{Cu$_{2}$OSeO$_{3}$}

\newcommand{\rxx}{$\rho_{xx}$}
\newcommand{\rxy}{$\rho_{xy}$}
\newcommand{\rxytop}{$\rho_{\rm xy}^{\rm top}$}
\newcommand{\Drxyt}{$\Delta\rho_{\rm xy}^{\rm top}$}
\newcommand{\Sxy}{$\sigma_{xy}$}
\newcommand{\Sxya}{$\sigma_{xy}^A$}

\newcommand{\bco}{$B_{\rm c1}$}
\newcommand{\bct}{$B_{\rm c2}$}
\newcommand{\bao}{$B_{\rm A1}$}
\newcommand{\bat}{$B_{\rm A2}$}
\newcommand{\beff}{$B_{\rm eff}$}
\newcommand{\bmm}{$B_{\rm m}$}
\newcommand{\bs}{$B_{\rm S}$}
\newcommand{\btop}{$B_{\rm top}$}
\newcommand{\bnfl}{$B_{\rm NFL}$}

\newcommand{\tc}{$T_{\rm c}$}
\newcommand{\tmax}{$T_{\rm max}$}
\newcommand{\ts}{$T_{\rm S}$}

\newcommand{\pc}{$p_{\rm c}$}

\newcommand{\mb}{$\mu_0\,M/B$}
\newcommand{\dmdb}{$\mu_0\,\mathrm{d}M/\mathrm{d}B$}
\newcommand{\ddmddb}{$\mathrm{\mu_0\Delta}M/\mathrm{\Delta}B$}
\newcommand{\cm}{$\chi_{\rm M}$}
\newcommand{\cac}{$\chi_{\rm ac}$}
\newcommand{\rechi}{${\rm Re}\,\chi_{\rm ac}$}
\newcommand{\imchi}{${\rm Im}\,\chi_{\rm ac}$}

\newcommand{\ozz}{$\langle100\rangle$}
\newcommand{\ooz}{$\langle110\rangle$}
\newcommand{\ooo}{$\langle111\rangle$}
\newcommand{\too}{$\langle211\rangle$}

\renewcommand{\floatpagefraction}{0.5}


\title{Supplementary Information:\\
Formation of a topological non-Fermi liquid in MnSi}

\author{R. Ritz}
\affiliation{Physik Department E21, Technische Universit\"at M\"unchen, D-85748 Garching, Germany}

\author{M. Halder}
\affiliation{Physik Department E21, Technische Universit\"at M\"unchen, D-85748 Garching, Germany}

\author{M. Wagner}
\affiliation{Physik Department E21, Technische Universit\"at M\"unchen, D-85748 Garching, Germany}

\author{C. Franz}
\affiliation{Physik Department E21, Technische Universit\"at M\"unchen, D-85748 Garching, Germany}

\author{A. Bauer}
\affiliation{Physik Department E21, Technische Universit\"at M\"unchen, D-85748 Garching, Germany}

\author{C. Pfleiderer}
\affiliation{Physik Department E21, Technische Universit\"at M\"unchen, D-85748 Garching, Germany}

\date{\today}


\maketitle

\section{Experimental methods}

A comprehensive account of the magneto-transport properties of MnSi under pressures up to $p^*\sim12\,{\rm kbar}$ has been given in Ref.\,\cite{Ritz:PRB2013}. This paper reports all the details of the experimental methods used, addressing in particular the importance of pressure transmitter, sample quality, as well as field and temperature history.

\section{Definition of transition fields and temperatures}

\begin{figure}
\begin{center}
\includegraphics[width=0.95\textwidth]{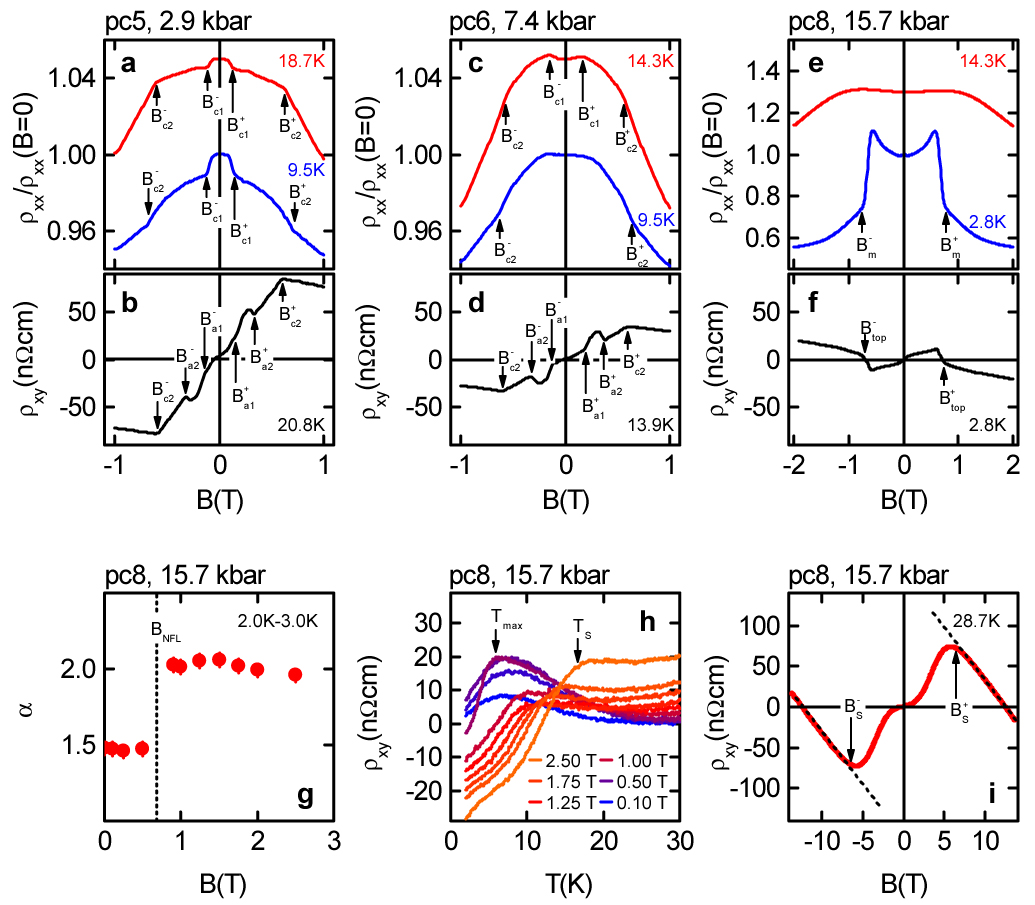}
\end{center}
\caption{Typical magneto-transport data illustrating key features used to define the various transition fields and temperatures. See text for further information.
\newline
\label{figureS1}
}
\end{figure}

Shown in Fig.\,\ref{figureS1} are typical data to illustrate characteristic features in the transport properties at the various phase transitions across the magnetic phase diagram and the definitions of the characteristic fields and temperatures used throughout the main text. The superscripts $+$ and $-$ address the behaviour under increasing and decreasing field, respectively. We note that data shown here for {\rxx} and {\rxy} represent sweeps from negative to positive fields calculated from a full field cycle in order to remove signal contributions from misaligned contacts. 

We begin with a low pressure of 2.9\,kbar shown Figs.\,\ref{figureS1}\,(a) and (b). Starting at zero field the magnetoresistance, {\rxx}, decreases for the field range shown here and changes slope at {\bco}, the helical to conical transition. This is followed by an additional change of slope at {\bct}, the conical to field-polarized ferromagnetic transition. At a temperature of 20.8\,K the Hall resistivity comprises three contributions: (i) the normal Hall effect, (ii) the anomalous Hall effect for which the Hall conductivity tracks the magnetization, and (iii) the topological Hall effect in the small field range of the A-phase. Accordingly at the conical to ferromagnetic transition at {\bct} the Hall resistivity displays a pronounced change of slope akin to that seen in the magnetization. In the A-phase an additional contribution to the Hall resisitivity is observed, limited by the transition fields {\bao} and {\bat}.  

The main features seen at low pressures are essentially unchanged at intermediate pressures as illustrated for 7.4\,kbar in Figs.\,\ref{figureS1}\,(c) and (d). A minor difference concerns the magnetoresistance in the helical state which increases slightly up to {\bco}. 

Figs.\,\ref{figureS1}\,(e) and (f) display the characteristic behaviour above the critical pressure $p_c$ for 15.7\,kbar. At low temperatures the magnetoresistance displays a distinct drop at the metamagnetic transition, denoted as $B_{\rm m}^+$ and $B_{\rm m}^-$. With increasing temperature the drop in {\rxx} evolves into a cross-over at increasing fields. The Hall resistivity is dominated by a topological Hall signal that extends all the way up to $B_{\rm top}^+$ and $B_{\rm top}^-$, which coincide with the metamagnetic transition fields. Due to the lower temperature, the intrinsic anomalous Hall contribution is very small; the change of {\rxx} at $B_{\rm m}^+$ and $B_{\rm m}^-$ does not account for the variation in {\rxy} in terms of an intrinsic anomalous Hall effect, which displays an entirely different temperature dependence.  This conclusion is underscored by the observation, that a similar topological Hall contribution is observed regardless of residual resistivity.

Shown in Fig.\,\ref{figureS1}\,(g) is a close-up view of the field dependence of the exponent $\alpha$ of the temperature dependence of the resistivity, as evaluated between 2\,K and 3\,K. The pronounced change from the $T^{3/2}$ non-Fermi liquid dependence to the $T^2$ Fermi liquid dependence illustrates the definition of $B_{\rm NFL}$. We refrain from showing data points in the immediate vicinity of $B_{\rm NFL}$.

\begin{figure}[h]
\begin{center}
\includegraphics[width=0.95\textwidth]{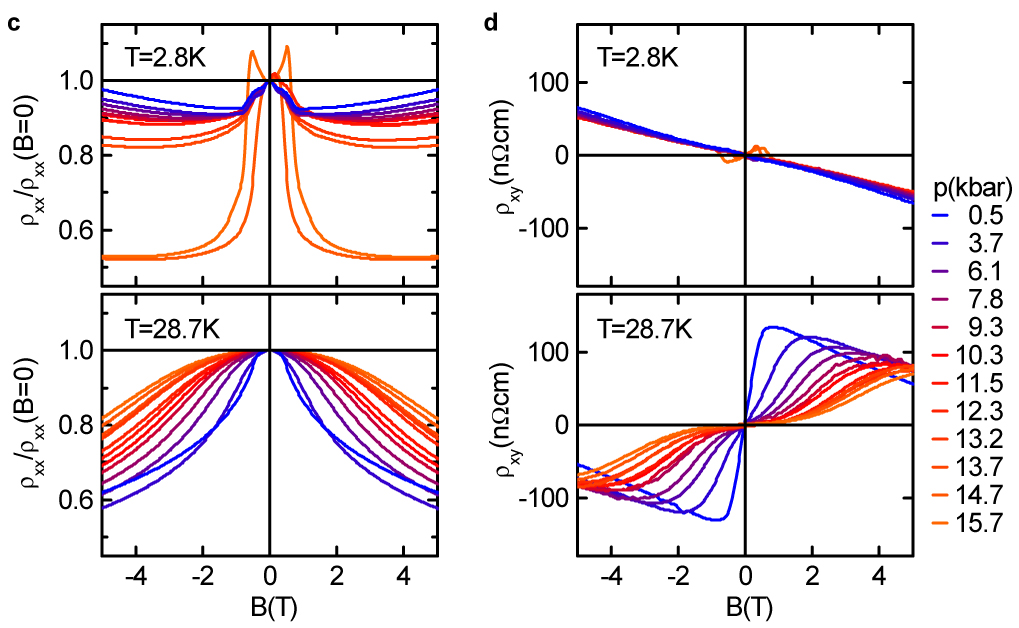}
\end{center}
\caption{Typical magneto-transport data at 15.7\,kbar (pressure cell pc8) providing an expanded view of Fig.\,1(c) and (d) of the main text.
\newline
\label{figureS1-2}
}
\end{figure}

The temperature dependence of the Hall signal for a wide range of magnetic fields is shown in Fig.\,\ref{figureS1}\,(h). For all magnetic fields the temperature dependence displays a decrease at low temperatures that is due to an intrinsic anomalous Hall contribution. Two field ranges may thereby be distinguished. At low magnetic fields a shallow maximum is observed at a temperature $T_{\rm max}$ that is clearly related to the temperature dependence of {\rxx}. Namely, the intrinsic anomalous Hall contribution, $\rho_{\rm xy}\propto M \rho^2_{\rm xx}$, decreases in good metals (like MnSi) with decreasing temperature, because the magnetization $M$ increases slower than the decrease in $\rho^2_{\rm xx}$. For larger fields (above $\sim1\,{\rm T}$), on the other hand, where {\rxx} decreases less strongly with decreasing temperature, the downturn of {\rxy} below a temperature $T_{\rm S}$ is related to the shoulder in the field dependence at $B_{\rm S}$ (see also Figs.\,\ref{figureS1}\,(i), \ref{figureS1-2}, \ref{figureS2}, and \ref{figureS2-2}). 

At sufficiently large pressures in the vicinity of $p_c$ the features at $B_{\rm S}$ and $T_{\rm S}$, respectively, connect with itinerant metamagnetism as observed in the ac susceptibility \cite{Thessieu:JPCM97}. However, as described in the main text, we find that the maximum at $T_{\rm S}$ cannot be accounted for by the temperature dependence of $M$ or {\rxx} or a simple combination of both. Fig.\,\ref{figureS1}\,(i) illustrates the definition of $B_{\rm S}$ associated with the S-shape in the Hall data for a high temperature of 28.7\,K deep in the paramagnetic regime (the pressure dependence is show in Fig.\,1\,(d) of the main text). The onset of the S-shape is denoted $B_{\rm s}^+$ and $B_{\rm s}^-$ as illustrated here.

\newpage
\section{Close-up view of data}

Here we presented Figs.\,1\,(c) and (d) of the main text on a smaller field scale to permit closer inspection of the detailed behaviour at low fields.

\newpage
\section{Hall-effect at high fields, temperatures, and pressures}

\begin{figure}[b]
\begin{center}
\includegraphics[width=0.5\textwidth]{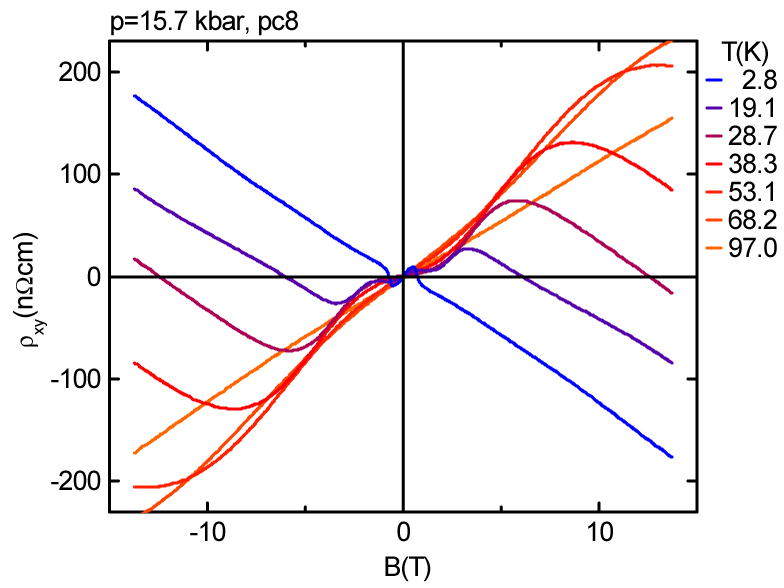}
\end{center}
\caption{Typical Hall resistivity above the critical pressure for 15.7\,kbar over a large field range of $\pm14\,{\rm T}$ and for various temperatures. With increasing temperature the anomalous Hall signal dominates, where an additional S-shape emerges with increasing pressure as illustrated in Fig.\,1\,(d) of the main text. The temperature dependence of the S-shape follows that of the intrinsic anomalous Hall effect. 
\label{figureS2}
}
\end{figure}

In this section we present additional data of the Hall resistivity over a wide field range of $\pm14\,{\rm T}$. A  characteristic of the Hall effect that emerges at high pressures and high temperatures concerns an additional S-shape as compared with the anomalous Hall signal expected from the magnetization (cf. Fig.\,\ref{figureS1}\,(i)). Associated with this S-shape is a reduction of the slope of the Hall resistivity at $B=0$. This was illustrated in Fig.\,1\,(d) of the main text as a function of pressure for a temperature of 28.7\,K. In comparison, shown in Fig.\,\ref{figureS2} is the evolution of the S-shape in the Hall resistivity up to 14\,T for a pressure of 15.7\,kbar as a function of temperature. At the lowest  and highest temperatures studied the anomalous Hall signal, together with the additional S-shape vanishes. 

\begin{figure}[h]
\begin{center}
\includegraphics[width=0.74\textwidth]{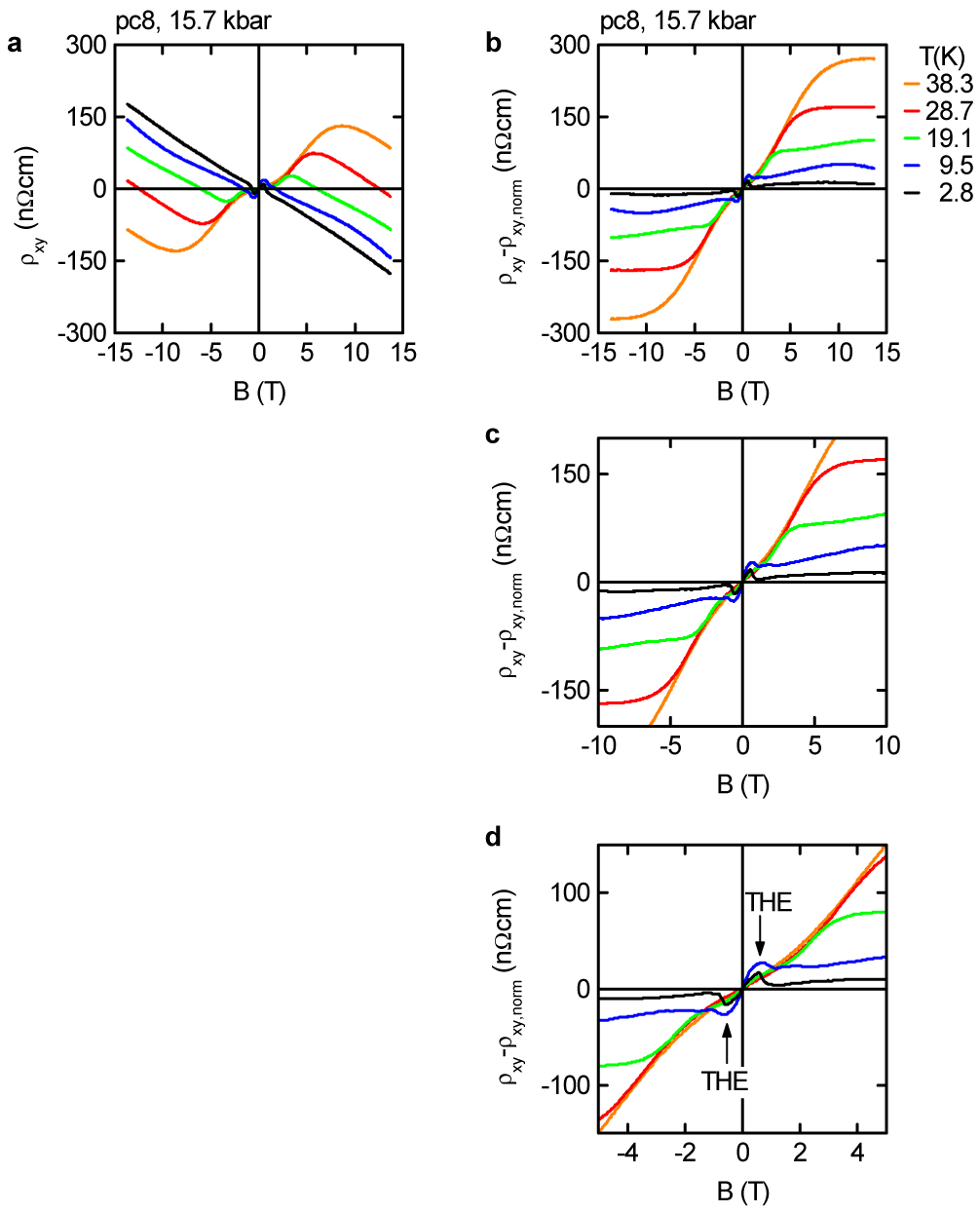}
\end{center}
\caption{Details of the temperature dependence of the Hall resistivity at 15.7\,kbar. (a) Data as shown in Fig.\,\ref{figureS2} for selected temperatures. (b) Data as shown in panel (a) after subtracting the normal Hall contribution, which is linear in field. The S-shape and reduced slope for $B=0$ are clearly visible. At high temperatures a shoulder is located at several Tesla that shifts to lower fields with decreasing temperatures and merges with an itinerant metamagnetic transition \cite{Thessieu:JPCM97}. The Hall signal at the largest fields is characteristic of an intrinsic anomalous contribution that \textit{decreases} with decreasing temperature. (c) \& (d) Data as shown in panel (a) on a smaller field scale. As the temperature decreases below $T^*$ (the temperature below which the NFL resistivity emerges) an additional maximum emerges, marked THE, which \textit{increases} with decreasing temperature. 
\label{figureS2-2}
}
\end{figure}

To illustrate that the S-shape is not the origin of the topological Hall signal in the NFL phase we show in Fig.\,\ref{figureS2-2}\,(a) selected low temperatures of Fig.\,\ref{figureS2} in further detail. After subtracting the same normal Hall contribution, which is linear in field, from all curves, the remaining Hall resistivity, $\rho_{\rm xy}-\rho_{\rm xy,norm}$, at high fields is characteristic of an intrinsic anomalous Hall effect (Fig.\,\ref{figureS2-2}\,(b)). Namely, it is essentially constant as a function of field and \textit{decreases} with decreasing temperature. As the temperature decreases the shoulder, which is located at several Tesla at high temperatures, moves towards lower fields and eventually merges with the itinerant metamagnetic transition inferred from ac susceptibility measurements \cite{Thessieu:JPCM97}.

Zooming now in on the behaviour on a smaller field scale, shown in Figs.\,\ref{figureS2-2}\,(b) \& (c), an additional maximum emerges below $\sim T^*$, i.e., in the range of the NFL resistivity, that \textit{increases} with decreasing temperature. Carefully tracking this signal to lower pressure this contribution, denoted THE, actually connects with the large topological Hall signal in the A-phase. The topological Hall signal appears thereby limited on the high field side (around one Tesla) by the shoulder in the anomalous Hall signal and thus the location of the itinerant metamagnetism \cite{Thessieu:JPCM97}. This consistency with the itinerant metamagnetism at high pressures and low temperatures empirically connects the emergence of the S-shape with the itinerant metamagnetism. 

As the NFL resistivity and topological Hall signal disappear for fields exceeding the itinerant metamagnetic transition, this suggests that the electronic structure and magnetic state change substantially at the itinerant metamagnetic transition. On the high field side conventional Fermi liquid behaviour is recovered. 

\newpage

\newpage
\bigskip
\section{Hall-effect in samples with low RRR}

\begin{figure}[h]
\begin{center}
\includegraphics[width=0.6\textwidth]{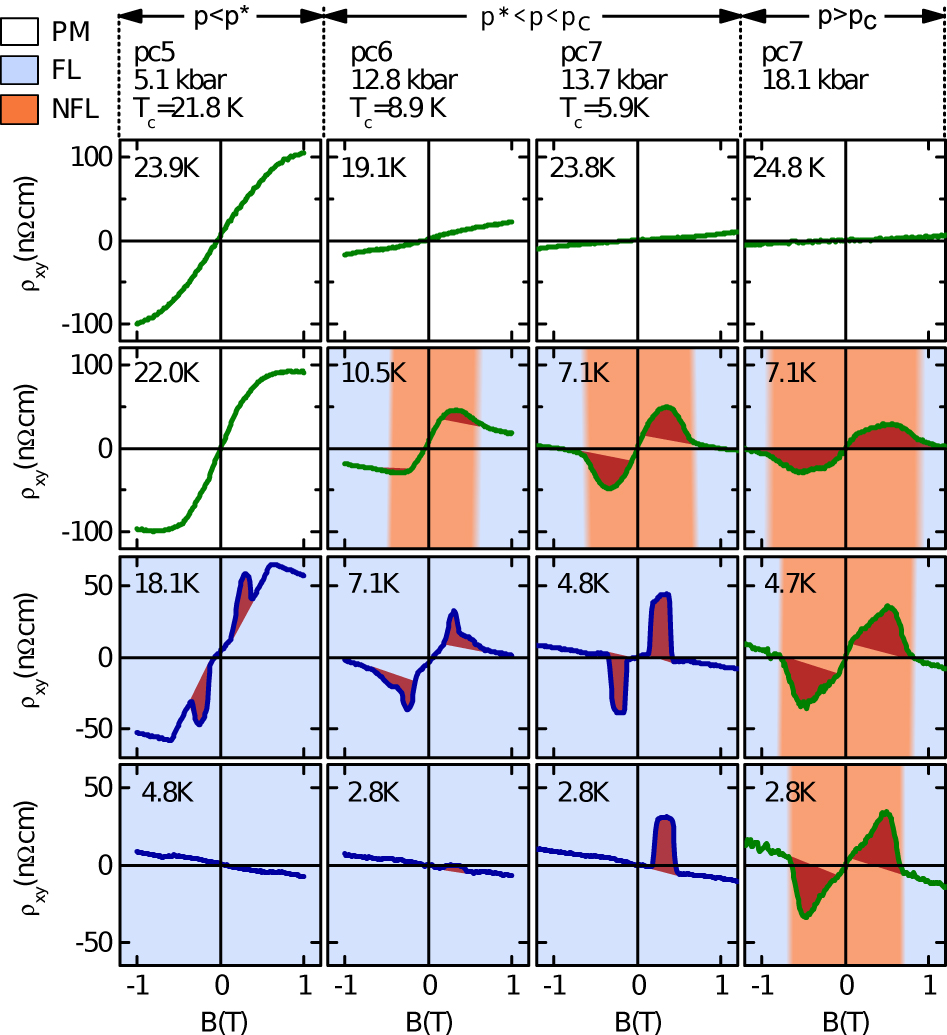}
\end{center}
\caption{Typical Hall resistivity for various temperatures and pressures for samples with low residual resistivity ratios (RRRs) around 45. This plot compares with Fig.\,2 of the main text where the same behaviour is reported for samples with high RRRs around 150.
\label{figureS3}
}
\end{figure}

Typical Hall resistivity for various temperatures and pressures in samples with low residual resistivity ratios (RRR) around $\sim 45$ is shown in Fig.\,\ref{figureS3}. This plot compares with the Hall data of a high RRR sample shown in Fig.\,2 of the main text. For the low RRR sample no qualitative differences are observed. In particular, this concerns the evolution of the topological Hall signal in the A-phase into a prominent characteristic of the NFL regime. However, the topological Hall signal is overall larger and the phase boundaries of the A-phase are less well defined. For pressures up to 12\,kbar data of this type was presented in great detail in Ref.\,\cite{Ritz:PRB2013}. We note that a strong hysteresis is observed as a function of field at 13.7\,kbar and 2.8\,K in the regime  where strong phase segregation was seen in neutron scattering \cite{Pfleiderer:Nature2004,Pfleiderer:Science07}, mu-SR \cite{Uemura:NaturePhysics07}, and NMR \cite{Yu:PRL2004}.

\newpage
\section{Importance of temperature and field history}

\begin{figure}[b]
\begin{center}
\includegraphics[width=0.7\textwidth]{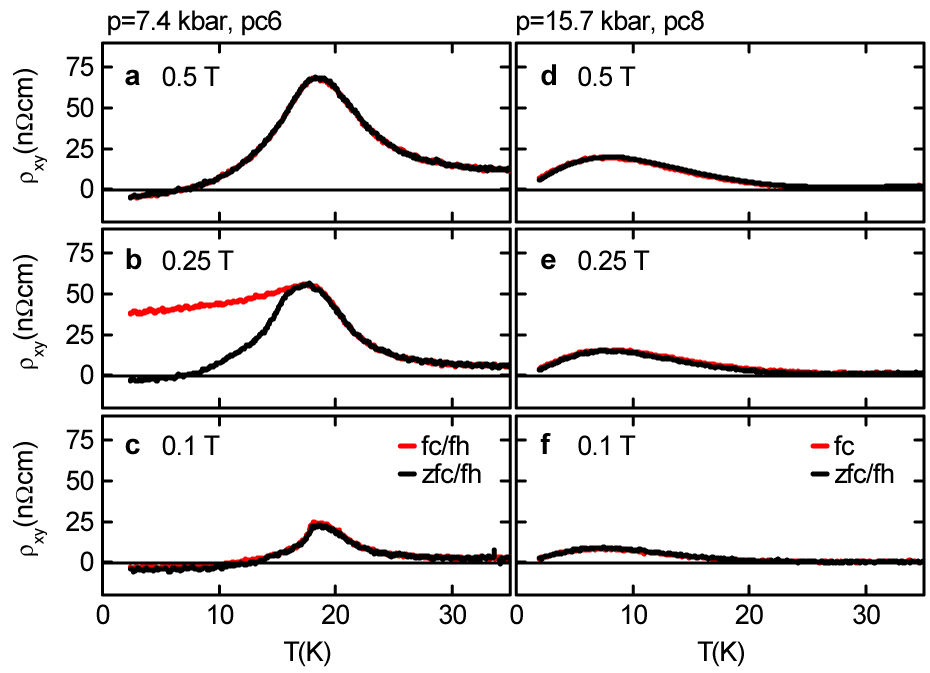}
\end{center}
\caption{
Temperature dependence of the Hall resistivity for magnetic field values of 0.5\,T, 0.25\,T, and 0.1\,T. For a field of 0.25\,T the temperature dependence crosses the skyrmion lattice phase. At intermediate pressures (here 7.4\,kbar) metastable behaviour is observed. Notably under field cooling (fc) the topological Hall contribution survives down to the lowest temperatures measured (zfc refers to zero-field cooling). No such metastable behaviour is observed above the critical pressure as shown here for 15.7\,kbar. For a detailed account of the metastable behaviour we refer to Ref.\,\cite{Ritz:PRB2013}.
\label{figureS4}
}
\end{figure}

For intermediate pressures the topological Hall signal is highly sensitive to the temperature and magnetic field history, where the size of the metastable signal depends slightly on the sample quality. The importance of this metastable behaviour has been discussed in great detail in Ref.\,\cite{Ritz:PRB2013}. An example illustrating the metastable behaviour for a pressure of 7.4\,kbar is shown in Fig.\,\ref{figureS4}, panels (a) through (c). For a field of 0.25\,T the topological Hall signal survives down to the lowest temperatures under field cooling (FC). 

In contrast, no metastable behaviour and no dependence on the field and temperature history is observed above the critical pressure, as illustrated in Fig.\,\ref{figureS4}, panels (d) through (f). Regardless of sample quality, the same Hall resistivity is observed under field cooling and zero-field cooling.

\newpage
\section{Summary of all pressure cells}

Table\,\ref{table:samples} summarizes key information of all pressure cells assembled for our study. The table represents an extended version of a similar table shown in Ref.\,\cite{Ritz:PRB2013}.

\begin{table*}[b]
\centering
\caption{Pressure cells prepared for our magneto-transport and magnetization measurements under pressure. Pressures are stated in the order in which they were applied. $^1$ The residual resistance ratio, RRR, was determined at the lowest pressure as the ratio $\rho_{xx} \left( T=280\,\rm{K} \right) / \rho_{xx} \left( T \rightarrow 0 \right)$. $^2$ FI: Fluorinert FC72:FC84 mixture at a 1:1 volume ratio. ME: methanol:ethanol mixture at a 4:1 volume ratio. In the transport measurements the electrical current was applied along the longest direction of the sample, the field was applied along the shortest direction.
\label{table:samples}}
\vspace{5mm}
\begin{tabular}{p{1.5cm}lp{2.6cm}lp{1.3cm}p{5cm}}
\hline\noalign{\smallskip}
\hline\noalign{\smallskip}
pressure cell (pc) & RRR$^1$ & sample size (${\rm mm^3}$) $l\times w\times t$& orientation & pressure medium$^2$ & pressures (kbar)\\
\hline\noalign{\smallskip}
1 & $\approx$93 & $2.9 \times 1.0 \times 0.22$ & $B \parallel t\parallel \langle 110 \rangle$, $I \parallel l\parallel \langle 100 \rangle$ & FI & 6.6 \\
2 & $\approx$92 & $2.8 \times 1.0 \times 0.25$ & $B \parallel t\parallel \langle 110 \rangle$, $I \parallel l\parallel \langle 100 \rangle$ & FI & 7.0, 10.0 \\
3 & $\approx$300 & $2.5 \times 1.0 \times 0.20$ & $B \parallel t\parallel \langle 110 \rangle$, $I \parallel l\parallel \langle 100 \rangle$ & ME & 10.7 \\
4 & $\approx$300 & $2.7 \times 0.9 \times 0.20$ & $B \parallel t\parallel \langle 110 \rangle$, $I \parallel l\parallel \langle 100 \rangle$ & ME & 8.1 \\
5 & $\approx$45 & $2.8 \times 1.0 \times 0.20$ & $B \parallel t\parallel \langle 110 \rangle$, $I \parallel l\parallel \langle 110 \rangle$ & ME & 7.6, 6.7, 5.9, 5.7, 5.1, 4.6, 3.4, 2.9, 2.6, 2.2, 0.3 \\
6 & $\approx$40 & $2.7 \times 1.0 \times 0.20$ & $B \parallel t\parallel \langle 110 \rangle$, $I \parallel l\parallel \langle 100 \rangle$ & ME & 5.2, 7.4, 9.6, 11.0, 11.2, 12.8, 0.4 \\
7 & $\approx$45 & $2.8 \times 1.0 \times 0.20$ & $B \parallel t\parallel \langle 110 \rangle$, $I \parallel l\parallel \langle 110 \rangle$ & ME & 18.1, 13.7 \\
8 & $\approx$150 & $2.9 \times 1.1 \times 0.20$ & $B \parallel t\parallel \langle 110 \rangle$, $I \parallel l\parallel \langle 100 \rangle$ & ME & 15.7, 14.7, 13.7, 13.2, 12.3, 11.5, 10.3, 9.3, 7.8, 6.1, 4.9, 3.7, 0.5\\
M & $\approx$70 & $6.0 \times 1.0 \times 1.0$ & $B \parallel l\parallel \langle 100 \rangle$ & ME & 0.0, 4.05, 7.50, 10.13, 11.80\\
\noalign{\smallskip}\hline
\noalign{\smallskip}\hline
\end{tabular}
\end{table*}


\newpage